\documentclass[11pt]{article}
\vspace{7cm} \textwidth 14cm \textheight 21.6cm
\usepackage{amsmath}
\usepackage{amsfonts}
\usepackage{url}
\usepackage[english]{babel}
\usepackage[dvipsnames]{xcolor}
\usepackage{hyperref}
\usepackage[pdftex]{graphicx}
\usepackage{caption}
\usepackage{subcaption}
\usepackage[utf8]{inputenc}
\usepackage{booktabs}
\usepackage{changes}
\usepackage{latexsym,amsthm,amscd}

\newcounter{alphthm}
\usepackage{tikz}
\usetikzlibrary{shapes.geometric, arrows}
\tikzstyle{startstop} = [rectangle, rounded corners, minimum width=1cm, minimum height=1cm,text centered, draw=black, fill=blue!30]
\tikzstyle{io} = [trapezium, trapezium left angle=70, trapezium right angle=110, minimum width=3cm, minimum height=1cm, text centered, draw=black, fill=blue!30]
\tikzstyle{process} = [rectangle, minimum width=5cm, minimum height=1cm, text centered, draw=black, fill=white!30]
\tikzstyle{decision} = [ellipse, minimum width=3cm, minimum height=1cm, text centered, draw=black, fill=white!30]
\tikzstyle{arrow} = [thick,->,>=stealth]
\tikzstyle{edge from parent}=[draw,dashed,thick,->,>=stealth,black]
\tikzstyle{process} = [rectangle, minimum width=3cm, minimum height=1cm, text centered, text width=3cm, draw=black, fill=white!30]
\theoremstyle{remark}
\usepackage{amssymb, graphicx, tcolorbox, wasysym,  pifont}
\numberwithin{equation}{section}

\newtheorem*{theorem*}{Theorem} 
\newtheorem{theorem}{Theorem}[section]

\theoremstyle{definition}

\newtheorem{example}[theorem]{Example}

\theoremstyle{remark}
\newtheorem{remark}[theorem]{Remark}

\title{An Approximated Model of Wildfire Propagation on Slope}

\author{ Hengameh R. Dehkordi \thanks{ORCID ID: 0000-0002-1738-3373, \hfill\break\indent
		Email: hengameh.r@ufabc.edu.br, \hfill\break\indent
		Phone: +55 (11) 4996-8332    	}\\
	  Center of Mathematics, Computation and Cognition, \\
	Federal University of ABC, \\
	 Santo André, Brazil}

\date{\today}
\date{}

\begin{document}
\maketitle


\begin{abstract}

The increasing frequency and intensity of wildfires underscore the need for accurate predictive models to enhance wildfire management. Traditional models, such as Rothermel and FARSITE, provide foundational insights but often oversimplify the complex dynamics of wildfire spread. Advanced methods, employing sophisticated mathematical techniques, offer more precise modeling by accounting for real-world complexities and dynamic environmental factors.

This paper focuses on wildfire propagation over inclined terrains and combines the Rothermel model, Huygens' principle, and advanced mathematical techniques to provide a more precise model of propagation. Environmental parameters and vegetation factors are directly incorporated into formulas and equations to improve the reliability and effectiveness of wildfire management strategies. The practical application of these results is demonstrated through MATLAB simulations, specifically examining wildfire spread under wind conditions that do not impede upwind fire advancement. The findings of this work contribute to both wildfire research and the development of more effective management strategies.
  
\end{abstract}


\section{Introduction}

Due to global warming, the frequency and intensity of wildfires increase annually. Wildfires threaten wildlife, forests, grasslands, and agricultural lands. Developing methods to study and predict fire behavior more accurately and reliably is crucial for effective wildfire management, as these methods can help mitigate the damage caused by fires. 

The Rothermel model is a mathematical model, widely utilized in wildfire management and research, developed in 1972 by Richard Rothermel to predict the spread rate of wildfires \cite{rothermel1972mathematical}. This model considers several factors, including vegetation characteristics, terrain slope, and wind speed. In the Rothermel model, vegetation is treated as a uniform and continuous fuel layer, simplifying the spatial heterogeneity found in real-world landscapes. Wind speed is assumed to be constant and its effect is incorporated through an empirical relationship that enhances the spread rate in the direction of the wind. For each specific area, the fire spread is often represented by an elliptical shape, where the major axis aligns with the wind direction, indicating faster spread rates in that direction. This simplification allows the model to predict fire spread more easily. Still, it may not fully capture the complexities of inhomogeneous fuel distributions and varying wind conditions observed in wildfires.

Another approach to modeling wildfire propagation involves using simulators, including Phoenix, IGNITE, Bushfire, FireMaster, FARSITE, and Prometheus \cite{Sullivan2009}. FARSITE (Fire Area Simulator) is one of the most commonly used wildfire simulators  \cite{farsite2004fire}. FARSITE integrates the Rothermel model as a fundamental component to simulate fire behavior and employs Huygens' principle for modeling fire spread. FARSITE assumes static fuel, weather, and topography conditions for each calculation, which may not fully capture the dynamic nature of real-world fire environments.

The Finsler geometry is a powerful tool to analyze real-world phenomena, in particular, wildfire propagations \cite{Markvorsen2016, markvorsen2017geodesic, Javaloyes2021, dehkordi2022applications, dehkordi2023applications}. Applying Finsler geometry to study wildfire propagation allows for analyzing and modeling more complex, real-world scenarios. Exploring wildfire propagation from the perspective of Finsler geometry is still in its nascent stages. First, Markvorsen showed the validity of Huygens' principle for Finsler spaces of dimension $2$ and used the elliptical frames to model the wildfire propagation \cite{Markvorsen2016}. Then, he used the so-called frozen metric, a time-dependent metric, to verify the wildfire propagation under a time-dependent wind \cite{markvorsen2017geodesic}. The authors in \cite{Javaloyes2021} verified the wildfire propagation with a strong wind by using the cone structure and then generalized the mode for the propagation on slope \cite{angel2023general}. Dehkordi in \cite{dehkordi2022applications} studied wildfire propagations in flat terrains under wind influence and provided wave equations to model the propagation. In \cite{dehkordi2023applications},  the concept of strategic paths was explored, demonstrating its application in wildfire management strategies to protect transmission lines against fire. Although many researchers have investigated the theoretical foundations of this problem, there exists a significant need for research that bridges the gap between traditional methods (e.g., the Rothermel model) and sophisticated mathematical techniques.

One novel contribution of this work is the combination of Finsler geometry techniques, the Rothermel model, and Huygens' principle to develop a more precise model of fire propagation on inclined terrain. The impact of slope on wildfire propagation, particularly in wildland areas, and the nonlinear relationship between slope and fire spread rates, as explored in \cite{dupuy2011slope}, underscores the need for further investigation. We model propagation by directly incorporating environmental and vegetation properties into formulas and equations. 
For the following four different scenarios which would happen in reality, more details are provided in Section 3:\\
Model 1: No wind, homogeneous environmental and vegetation properties.\\
Model 2: Wind present, homogeneous vegetation properties.\\
Model 3: No wind, inhomogeneous vegetation properties.\\
Model 4: Wind present, inhomogeneous vegetation properties.\\
Windless models are more suitable for laboratory and experimental settings, while models with homogeneous vegetation distribution represent scenarios where a specific vegetation type dominates an area. All models provide equations for fire location, spherical wave propagation, and fire spread paths. The latter is crucial for identifying strategic locations for firebreaks, and addressing the blocking problem (see Bressan \cite{bressan2013} and references therein). Additionally, spherical wave equations enable the application of Huygens' principle to determine fire front propagation.
This work paves the way for interdisciplinary collaborations between mathematics, engineering, natural sciences, and risk management.

Three principal factors for a wildfire to spread are radiation, conduction, and convection \cite{andrews2018rothermel}. Radiation is the heat transfer to the nearby fuel, convection is the movement of hot air upwards, and conduction is the heat transfer through direct contact between vegetation. Radiation and convection are the primary heat transfer mechanisms in wildfire propagation, while conduction plays a secondary role and is often simplified in wildfire propagation models \cite{speer2022wildland}. The Rothermel model primarily incorporates the radiation mechanism to describe the heat transfer process. This model does not explicitly model convection and conduction to the same depth as radiation. Convection is considered indirectly, mainly through its effect on preheating fuels and influencing the spread rate. 
We study wildfire propagation on slope by analyzing radiation and convection. The fundamental idea is to use a Matsumoto-type Finsler metric \cite{matsumoto1989slope}.
 We show that, without wind, the propagation wave from a point is the sum of two circles.   With wind, it is the sum of an ellipse and a circle. In the literature, an elliptical frame is often used to model propagation even when the terrain is inclined; see, for instance, \cite{rothermel1972mathematical, anderson1982modelling}. In all the cases, a Matsumoto or a modified Matsumoto metric corresponds to the problem.

Wind is important in wildfire propagation because it increases the spread rate by supplying oxygen to the fire. Additionally, wind can change the direction of the propagation, making it progress downwind. In certain scenarios, especially when the wind direction opposes the fire’s path, strong winds can potentially slow down or hinder the spread of the fire; however, in most cases, wind speeds are generally not strong enough to prevent the fire from advancing upwind. Due to the high potential of application, this paper focuses on wildfire spreading under mild wind conditions, where the wind does not obstruct the fire from spreading upwind. When the wind impedes the fire from propagating upwind, the center of the frame is not located inside of it. Therefore, a conic metric and more complex mathematics (see \cite{angel2023general}) are required to verify these cases. Since these scenarios are infrequent, they are not addressed in this work.

In practical scenarios, wind speed and direction can vary over time. In highly dynamic or turbulent conditions, such as near obstacles or during storm events, wind direction may change rapidly, on the order of seconds to minutes. Conversely, in more stable conditions, influenced by large-scale pressure systems or consistent weather patterns, wind direction and speed typically remain relatively stable for longer periods, ranging from minutes to hours, before undergoing transition. For detailed analyses of these phenomena, see for example \cite{mcsweeney2019temporal}, \cite{lopez2022analysis}, and \cite{dai1999diurnal}.
Assuming constant wind speed and direction over specific periods in wildfire modeling is practical and beneficial. This assumption aligns with real-world scenarios and simplifies the associated metrics and mathematical tools, making them time-independent. Consequently, this assumption reduces computational complexity and minimizes errors. This work assumes that the wind remains time-independent within a time interval and transitions to a new time-independent wind in the subsequent interval. When wind conditions change, the problem shifts to a new propagation scenario, which is analyzed by considering the perimeter of the burnt area as the initial wave of the new propagation problem.


The rest of this work is organized as follows. In the next section, we present some preliminary concepts. Section 3 states the main results through two subsections, which verify the homogeneous and nonhomogeneous distribution of vegetation characteristics.
To present the models, section 3 considers two cases of propagation in the absence of wind ({Models 1} and {2}) and two cases in the presence of wind ({Models 3} and {4}). The wind may be either space-dependent or space-independent.
Section 4 includes two flowcharts that illustrate the process of modeling propagation and deriving the equations for fire fronts and paths, facilitating the application of the results.
Section 5 presents some examples that demonstrate the results' applications. Finally, Section 6 is dedicated to the conclusion and final remarks.


\section{Preliminaries}

A Finsler metric on the plane $M : z = d_1x + d_2y + d$, is a function $F: TM \to [0, \infty)$, where $TM $ is the set of all points $p=(x,y)$ of $M$ and tangent vectors at $p$, $T_pM$. The metric $F$ must be positive homogeneous, meaning that for any vector $v \in T_pM$ and any scalar $\lambda > 0$, $F(p, \lambda v) = \lambda F(p, v)$. Furthermore, $F$ must be smooth on the tangent bundle $TM$, excluding the zero section. Additionally, $F$ must be positively definite, implying its fundamental form, 
\begin{equation}\label{sff}
     g_{ij} = \frac{1}{2} \frac{\partial^2 F^2}{\partial v^i \partial v^j}, \ i,j=1,2,
\end{equation}
is positive definite. The Finsler metric $F$ assigns a length to each tangent vector $v=(v_1,v_2)$ in a way that depends smoothly on both the base point $p$ and the vector $v$. Let $\gamma : [a, b] \rightarrow M$ be a piecewise smooth curve in $M$. The Finsler functional $\mathcal{L}$ associated with this curve is defined as:
\[
\mathcal{L}[\gamma] = \int_a^b F(\gamma(t), \dot{\gamma}(t)) \, dt.
\]
The curve $\gamma(t)$ is a geodesic if it locally minimizes the Finsler length functional. One shows that the geodesics are the solutions to the following system of second-order differential equations:
\begin{equation}\label{geo-fin}
    \frac{d^2 x^i}{dt^2} + 2 G^i\left(x, \frac{dx}{dt}\right) = 0,  \ i=1,2,
\end{equation}
where 
\[
G^i(x, v) = \frac{1}{4} g^{il} \left( 2 \frac{\partial g_{jl}}{\partial x^k} v^j v^k - \frac{\partial g_{jk}}{\partial x^l} v^j v^k \right),
\]
where $g_{ij}$ is the fundamental form, given in Eq.~\eqref{sff}, and $g^{ij}$ is its inverse.


By a fire front at time $t$, we mean the perimeter of the burnt area at $t$, and fire rays are the paths along which the fire advances. The fire front is spherical if the fire starts from a single point. For each wildfire propagation problem, we define a frame as the perimeter of the area burned by the fire, which starts from an ignition point and spreads across the terrain. At the same time, the wind and vegetation characteristics are assumed to be consistent with those at the ignition point. We use this frame to determine the Finsler metric.

The following theorem forms the basis for finding the propagation models in this work. In this theorem, $A$ represents the subset of $M$ whose perimeter constitutes the fire front, and $U$ denotes the area affected by the fire during the time of study.

\begin{theorem}\label{one} \cite{dehkordi2019huygens}
Assume $(M, F)$ is a Finsler space across which a wildfire spreads and $A$ is a compact subset of $M$. We define $\rho: M \to \mathbb{R}$ by $\rho(p) = d_F(A, p)$ and assume $\rho^{-1}([s, r]) = U$ and there is no cut locus in $\rho^{-1}(s, r)$. If $\rho^{-1}(s)$ is the fire front at time $0$, then, $\rho^{-1}(t)$ is the fire front at time $t - s$, $t \in [s, r]$, and Huygens' principle is satisfied by the fire fronts $\{\rho^{-1}(t)\}_{t \in [s, r]}$. Moreover, each fire ray is a geodesic orthogonal to each fire front $\rho^{-1}(t)$ at time $t - s$.
\end{theorem}

Theorem \ref{one} asserts that if a Finsler metric is identified such that the perimeter of the Finsler distance function coincides with the fire front, then this distance function models the fire propagation. Furthermore, the level sets of this distance function satisfy Huygens' principle. This principle is named for the 17th century Dutch mathematician Christian Huygens who proposed it for describing the travel of light waves. We state Huygens' principle as follows:

\begin{theorem}\cite{arnol2013mathematical}
	Let $\phi_p(t)$ be the spherical wavefront of the point $p$ after time $t$. For every point $q\in\phi_p(t)$, consider the spherical wave front after time $s$, i.e. $\phi_q(s)$. Then, the spherical wavefront of point $p$ after $s+t$ will be the envelope of spherical wavefronts $\phi_q(s)$.
\end{theorem}

In the next section, we verify Models 1-4 mentioned in the Introduction section, validate them,  and provide the fire fronts and rays equations.

\section{Propagation on an Inclined Plane}
This section models wildfire propagation on an inclined plane under two situations: homogeneous and nonhomogeneous distribution of vegetation characteristics. Throughout this section, \(\langle \cdot , \cdot \rangle\) denotes the standard Euclidean inner product.

It is noteworthy that in reality, the total area affected by a wildfire consists of subareas where the vegetation can be considered homogeneous, and the wind might remain constant during a time interval. Therefore, one can apply the simpler models, including {Models 1 } and {2}, for subareas with homogeneous vegetation characteristics, and use {Models} {3} and {4} for more complex vegetation characteristics. By doing so, we avoid errors that arise from complex computations and provide a more reliable model by studying each subarea's characteristics as the fire reaches it.

\subsection{Homogeneous Distribution of Vegetation Characteristics}
Each specific geographic area may contain subregions with homogeneous vegetation characteristics—such as type, density, and moisture content. The agricultural lands are special cases; for instance, reference \cite{pagadala2024measuring}. This homogeneity in vegetation properties facilitates the development of simplified mathematical models for wildfire propagation. The homogeneous condition assumption aligns with many real-world scenarios where detailed or dynamic data may be unavailable or unnecessary for certain analyses. Any variation in conditions introduces a new propagation problem.

\vspace{5mm}
\textbf{Model 1.}
              We assume that a wildfire propagates on the inclined plane $M$, with no wind, the vegetation characteristic distribution being homogeneous, and the initial fire front being $A$.
Then, the fire rays are line segments $\gamma(t)=p+tv$, where $p$ are points of $A$ and $v$ are vectors such that 
\begin{equation}\label{alp1}
{||v||^2} - R_0(||v|| + \phi_s v_1) = 0
\end{equation}
and  
\begin{equation}\label{alp12}
\langle v, u \rangle - R_0\left(\frac{\langle v, u \rangle}{||v||} + \phi_s u_1\right) = 0,
\end{equation}
where $u = (u_1, u_2)$ is any vector tangent to $A$, $\phi_s$ is the slope factor and $R_0(1 + \phi_s)$ is the spread velocity towards the uphill. The parameters $\phi_s$ and $R_0$ are determined using data and tables \ref{t1} and \ref{t2} of Appendix \ref{apen3} and we must have $\phi_s<\frac12$ to apply the model. \\
Furthermore, the fire front at each time \( T \) is obtained using any of the following methods.
\begin{itemize}
    \item[(i)] Finding the set $\{\gamma(T)\ |\ \gamma(t)\ \text{ are fire rays from } A\}$.       \item[(ii)] Applying Huygens' principle to $A$, where the spherical fire fronts are given by Eq.~\eqref{ind} below:
    \begin{equation}\label{ind}
    r = R_0(1 + \phi_s \cos \theta).
    \end{equation} 
   \end{itemize}

\vspace{5mm}\textit{Validation of Model 1:} To validate the model, we apply the same idea of Matsumoto for the walker on a slope.
When a fire starts, radiation and convection are the principal factors that contribute to wildfire spread. Radiation plays the primary role in the absence of wind, causing the fire to spread in a circular shape with a radius \( R_0 \) from the ignition point. The value of \( R_0 \) is calculated using Appendix \ref{apen3}, tables \ref{t1} and \ref{t2}, and the information on the vegetation characteristics, considering both zero slope and zero wind.

 On a slope, both convection and radiation influence fire propagation. Convection currents still move upward but have a greater impact uphill than downhill. Therefore, the vector \(\Vec{c}\cos\theta\), where \(\theta\) is the angle relative to the upslope and \(\Vec{c}\) is the projection of the hot air current vector on the slope toward the uphill, influences fire propagation on the slope. Consequently, the fire front forms the following curve:

\begin{equation*}
r=R_0+c\cos\theta,
\end{equation*}
 where $R_0+c$ is the velocity of fire in the upslope direction. According to the Rothermel model \cite{andrews2018rothermel} (see pages 87-88), the velocity of fire in the upslope direction is $R_0(1+\phi_s)$ and, therefore, we have
\begin{equation}\label{limm1}
r=R_0(1+\phi_s\cos\theta), 
\end{equation}
where $\phi_s$ is determined by Appendix \ref{apen3}. 

We first find the Finsler metric associated with the problem to provide the propagation waves. This metric satisfies $F(v)=1$, where $v=(v_1,v_2)$ are vectors such that $||v||=r$ satisfies Eq.~\eqref{limm1}, with $||.||$ being the standard Euclidean norm. We replace $||v||=r$ in Eq.~\eqref{limm1} and apply Okubo's technique \cite{shen2001lectures}, which involves substituting $v\mapsto \frac{v}{F(v)}$ and finding $F(v)$, and obtain

\begin{equation}\label{fins1}
    F(v)=\frac{||v||}{R_0(1+\phi_s\cos\theta)},
\end{equation}
where $\theta$ is the direction of $v$ relative to the uphill.  One can verify that the inner product associated with the metric is positive definite if $\phi_s<\frac12$.

Next, we consider the Finsler distance function $\rho:M\longrightarrow \mathbb{R}$; $\rho(p)=d_F(A,p)$. Therefore, we have $\rho^{-1}(0)=A$ and apply Theorem \ref{one}. According to Theorem \ref{one}, the geodesics of the Finsler metric $F$, Eq.~\eqref{fins1}, provide the fire rays and fire front equations. Since the metric Eq.~\eqref{fins1} has no dependency on $(x,y)$, the geodesic equation \ref{geo-fin} becomes 
\begin{equation}\label{ge}
    (\frac{d^2x}{dt^2},\frac{d^2y}{dt^2})=(0,0),
\end{equation}
implying that the geodesics are straight lines. Therefore, by Theorem \ref{one}, given each point $p$ of $A$, the fire ray originating from $p$ is $\gamma(t)=p+tv$ where $v=(v_1,v_2)$ is a unit vector orthogonal to $A$. It is straightforward to see that the vector $v$ being unitary is equivalent to
\[
{||v||^2}={R_0(||v||+\phi_sv_1)}
\]
and being orthogonal to $A$ is equivalent to
\begin{equation*}
    <v,u>=R_0\left(\frac{<v,u>}{||v||}+\phi_su_1\right),
\end{equation*}
where $u=(u_1,u_2)$ is any vector tangent to $A$ at $p$.

To find the fire fronts, by Theorem \ref{one}, $\rho^{-1}(T)$ is the fire front at each time $T$. Moreover, the fire rays originating from $A$, $\{\gamma(t)\}$, are unitary geodesics and orthogonal to $A$; therefore, they simultaneously reach $\rho^{-1}(T)$. This implies that the set $\{\gamma(T)\}$ provides the fire front at $T$, confirming assertion (i) in the model.

To show assertion (ii), according to Theorem \ref{one}, Huygens' principle is satisfied by the fire front provided by the set $\{\rho^{-1}(t)\}$. Furthermore, the fire front given by Eq.~\eqref{limm1} is the Finsler geometric sphere as it coincides with $\{v\ |\ F(v)=1\}$ and the tangent space coincides with the plane $M$. This concludes item (ii) and completes the validation.

The next model studies the case in which the vegetation characteristics are homogeneous across the space and constant wind blows.

\vspace{5mm}\textbf{Model 2.
}    We assume that a wildfire propagates on the inclined space $M$, the vegetation characteristic distribution is homogeneous,  the initial fire front is $A$, and the wind blows. Then, the fire rays are $\gamma(t)=p+tv$, where $p\in A$ and $v$ are vectors that satisfy the following conditions:

    \begin{equation}\label{con1}
    ||v||^2=\frac{ab||v||^2}{\sqrt{(a^2-b^2)<v,\mathbf{v}_{\hat{\theta}}>^2+b^2||v||^2}}+c<v,\mathbf{v}_{\hat{\theta}}>,
\end{equation}

\begin{equation}\label{con2}
    \frac{(1-ab)<v,u>}{||v||^2}+\frac{(a^2-b^2)<v,\mathbf{v}_{\hat{\theta}}><u,\mathbf{v}_{\hat{\theta}}>+b^2<v,u>}{ab\sqrt{(a^2-b^2)<v,\mathbf{v}_{\hat{\theta}}>^2+b^2||v||^2}}+\frac{c<u,\mathbf{v}_{\hat{\theta}}>}{ab}=0,
\end{equation}
where $\mathbf{v}_{\hat{\theta}} = (\cos \hat{\theta}, \sin \hat{\theta})$ and $\hat{\theta}$ is the wind direction concerning upslope, $a=\frac{(1+0.25U)}{2(R_B+R_H)}$, $b=\frac12(R_B+R_H)$, $c=\frac12(R_H-R_B)$, and $u=(u_1,u_2)$ is any vector tangent to $A$ at $p$. Here $U$ is the speed of the wind in mid-flame, $R_H$ is the downwind rate of spread, and $R_B$ is the upwind rate of spread, all determined by data and tables \ref{t1} and \ref{t2} of Appendix \ref{apen3} and we must have $R_H^2<\frac98(1+0.25U)$ to be able to apply the model. 

Furthermore, the fire front at each time \( T \) can be obtained using any of the following methods.\begin{itemize}
    \item[(i)] Finding the set $\{\gamma(T)\ |\ \gamma(t)\ \text{\ are the fire rays from}\ A\}$.
 \item[(ii)] Applying Huygens' principle in which the spherical fire front is 
    \begin{equation}\label{fram2}
     r=\frac{ab}{\sqrt{a^2\cos^2(\theta-\hat{\theta})+b^2\sin^2(\theta-\hat{\theta})}}+c\cos(\theta-\hat{\theta}).
    \end{equation}
   \end{itemize}
   
\vspace{5mm}\textit{Validation of Model 2:}    When the wind blows on an inclined plane and the vegetation distribution is homogeneous, in the initial stages the fire shape is primarily influenced by wind, the projection of the heat vector on the slope, and fuel conditions. Therefore, a closed curve including a translated ellipse, egg shape, oval, and tear shape can approximate the fire shape (see the figures in \cite{hilton2015effects, glasa2011note}). To approximate the fire shape in polar coordinates, we use the following equation:
    \begin{equation}\label{fram3}
        r=\frac{ab}{\sqrt{a^2\cos^2(\theta-\hat{\theta})+b^2\sin^2(\theta-\hat{\theta})}}+{c}\cos(\theta-\hat{\theta}),
    \end{equation}
    where $b+c$ is the fire speed downwind, $b-c$ is the fire speed upwind, $a$ is the the the flanking fire speed, and $\hat{\theta}$ is the wind direction concerning upslope. Here, we consider the direction of the fastest speed toward the wind direction, see figures in \cite{viegas2004slope}. By Rothermel model, pages 87-88 of \cite{andrews2018rothermel}, we approximate $b=\frac12(R_H+R_B)$, $c=\frac12(R_H-R_B)$, and $a=\frac{1+0.25U}{2(R_H+R_B)}$, all determined by data and tables \ref{t1} and \ref{t2} of Appendix \ref{apen3}. It is not difficult to 
 see that Eq.~\ref{fram3} is convex if $2c<b<a+c$ and it can yield a variety of shapes, including ellipse, egg shape, and limaçon. One verifies that $2c<b<a+c$ if $\frac{R_H}{R_B}<3$ and $R_H^2<\frac98(1+0.25U)$. The first condition is always satisfied and we must check the later. 
    
     We follow the same process as that of Model 1 to find the Finsler metric, fire rays, and fire fronts. We consider the vectors $v=(v_1,v_2)$ such that $||v||=r$ and write the equation \ref{fram3} as
     \begin{equation}\label{fram4}
         ||v||=\frac{ab}{\sqrt{(a^2-b^2)\cos^2(\theta-\hat{\theta})+b^2}}+c\cos(\theta-\hat{\theta}).
    \end{equation}
     By substituting $v\mapsto \frac{v}{F(v)}$ into Eq.~\eqref{fram4}, we find the metric $F(v)$ as
    \begin{equation}\label{fins2}
        F(v)=\frac{||v||}{\frac{ab}{\sqrt{(a^2-b^2)\cos^2(\theta-\hat{\theta})+b^2}}+c\cos(\theta-\hat{\theta})}, 
    \end{equation}
    where $\theta$ is the direction of $v$ with the uphill.

To obtain the fire rays, we first verify the conditions equivalent to being unitary and orthogonal. By some straightforward calculations, the vector $v$ being unitary is equivalent to 
\begin{equation}\label{uni2}
    ||v||^2=\frac{ab||v||^2}{\sqrt{(a^2-b^2)<v,\mathbf{v}_{\hat{\theta}}>^2+b^2||v||^2}}+c<v,\mathbf{v}_{\hat{\theta}}>,
\end{equation}
where $\mathbf{v}_{\hat{\theta}} = (\cos \hat{\theta}, \sin \hat{\theta})$
and being orthogonal to $A$ is equivalent to 
\begin{equation}\label{orto2}
    \frac{(1-ab)<v,u>}{||v||^2}+\frac{(a^2-b^2)<v,\mathbf{v}_{\hat{\theta}}><u,\mathbf{v}_{\hat{\theta}}>+b^2<v,u>}{ab\sqrt{(a^2-b^2)<v,\mathbf{v}_{\hat{\theta}}>^2+b^2||v||^2}}+\frac{c<u,\mathbf{v}_{\hat{\theta}}>}{ab}=0,
\end{equation}
where  
$u=(u_1,u_2)$ is every vector tangent to $A$ at $p$.

By the same reasoning as in Model 1, the wave rays are straight line segments \(\gamma(t)=p+tv\), where \(p \in A\) and \(v\) are vectors that satisfy the conditions of Eqs.~\eqref{uni2} and \eqref{orto2}. Additionally, the fire front at each time \(T\) is \(\{\gamma(T)\}\). Applying Huygens' principle to the fire front \(A\), one can determine the subsequent fire fronts, with the spherical fire front given by Eq.~\eqref{fram3}.

\begin{remark}
In Models 1 and 2, if \(A\) is a point, then the fire ray originating from \(A\) is \(\gamma(t)=tv\), where \(v\) satisfies Eq.~\eqref{alp1} and Eq.~\eqref{con1}, respectively.
\end{remark}


\subsection{Nonhomogeneous distribution of vegetation characteristics}

In this section, we address cases where the vegetation characteristics are not homogeneous and vary smoothly across the space. For scenarios involving conditions with a finite number of non-smoothness or areas with accumulated non-smoothness conditions, we can approximate the model, employing {Models 3} and {4} given below that respectively discuss the no wind and with wind scenarios. These models cover more general propagation cases and the fire rays are typically not straight lines.

\vspace{5mm}\textbf{Model 3.} We assume that a wildfire propagates on the inclined plane \( M \), with no wind, the vegetation characteristic distribution varying smoothly, and the initial fire front being \( A \).
Then, the fire rays are solutions \( \gamma(t)=(x(t),y(t)) \) of the system of equations \eqref{geo-fin} with the metric given by Eq.~\eqref{fins3}, such that \( \gamma(0)=p\in A \) and \( \gamma'(0)=v \) are vectors satisfying
\begin{equation}\label{alp2}
{||v||^2} - R_0(p)(||v|| + \phi_s(p) v_1) = 0
\end{equation}
and  
\begin{equation}\label{alp3}
\langle v, u \rangle - R_0(p)\left(\frac{\langle v, u \rangle}{||v||} + \phi_s(p) u_1\right) = 0,
\end{equation}
where \( u = (u_1, u_2) \) is any vector tangent to \( A \), and at each point \( p \), \( \phi_s(p) \) and \( R_0(p)(1 + \phi_s(p)) \) are slope factor and the spread velocity towards the uphill, respectively, and determined using data and tables \ref{t1} and \ref{t2} of Appendix \ref{apen3}. The model is valid if $\phi_s<\frac12$.\\
Furthermore, one finds the fire front at each time \( T \) using each of the methods below.
\begin{itemize}
    \item[(i)] Finding the set \( \{\gamma(T)\ |\ \gamma(t)\ \text{ are fire rays from } A\} \).
    \item[(ii)] Applying Huygens' principle to \( A \), where at each point \( p \in A \), the spherical fire front is \( \{\gamma(1)\} \) such that \( \gamma(0)=p \) and \( ||v|| = R_0(p)(1 + \phi_s(p) \cos \theta) \). 
\end{itemize}

\vspace{5mm}\textit{Validation of Model 3:} Given any point $ p$ on $ M$, we assume that the tangent space $ T_pM$ shares the same characteristic vegetation properties as $ p$. This assumption allows us to apply {Model 1} and analyze propagation within $ T_pM$, determining the values of $ R_0$ and $ \phi_s$, and the associated metric. Since vegetation characteristics vary smoothly across $ M$, we can define smooth functions $ R_0$ and $ \phi_s$ over $ M$ such that at each point $ p$, these functions align with the local values of $ R_0$ and $ \phi_s$ specific to that point. Consequently, the Finsler metric on $ M$ resembles the Eq.~\eqref{fins1}, with $ R_0$ and $ \phi_s$ as smooth functions, leading to the conclusion that the Finsler metric associated with the propagation is \begin{equation}\label{fins3}
    F(p,v)=\frac{||v||}{R_0(p)(1+\phi_s(p)\cos\theta)},
\end{equation}
where \( \theta \) is the direction of \( v \) relative to the uphill. The inner product associated with the metric \ref{fins3} is positive definite if $\phi_s(p)<\frac12$.

We provide a similar approach as in {Model 1} to validate the model. We define the Finsler distance function \( \rho : M \to \mathbb{R} \) by \( \rho(p) = d_F(A, p) \) and apply Theorem \ref{one} to model the propagation. The unitary Finsler geodesics \( \gamma(t) \) that are orthogonal to \( A \) represent the fire rays. This involves analyzing the geodesic system \eqref{geo-fin} with initial conditions \( \gamma(0) \in A \), \( F(\gamma'(0)) = 1 \), and \( \gamma'(0) \) being \( F \)-orthogonal to \( A \). Through some calculations, we show that these unit and orthogonality conditions are equivalent to Eqs.~\eqref{alp2} and \eqref{alp3}. One verifies Assertions (i) and (ii) using similar arguments as in {Model 1}.

If the vegetation characteristics vary smoothly across the space and wind is present, we can establish the following model.

\vspace{5mm}\textbf{Model 4.} We assume that a wildfire propagates on the inclined plane $M$, the wind blows, the vegetation characteristic distribution varies smoothly, and the initial fire front is $A$.
Then, the fire rays $\gamma(t)=(x(t),y(t))$ are solutions of equations system \eqref{geo-fin}, such that $\gamma(0)=p\in A$ and $\gamma'(0)=v$ are vectors satisfying
 \begin{equation}\label{sph}
    ||v||^2=(\frac{ab||v||^2}{\sqrt{(a^2-b^2)<v,\mathbf{v}_{\hat{\theta}}>^2+b^2||v||^2}}+c<v,\mathbf{v}_{\hat{\theta}}>)(p),
\end{equation}
where $\mathbf{v}_{\hat{\theta}(p)} = (\cos \hat{\theta}(p), \sin \hat{\theta}(p))$, with $ \hat{\theta}(p)$ being the wind direction with upslope at $p$, and

\begin{equation}\label{sph2}
    (\frac{(ab-a^2b^2)<v,u>}{||v||^2}+\frac{(a^2-b^2)<v,\mathbf{v}_{\hat{\theta}}><u,\mathbf{v}_{\hat{\theta}}>+b^2<v,u>}{ \sqrt{(a^2-b^2)<v,\mathbf{v}_{\hat{\theta}}>^2+b^2||v||^2}}+{c<u,\mathbf{v}_{\hat{\theta}}>})(p)=0,
\end{equation}
where  
$u=(u_1,u_2)$ is every vector tangent to $A$ at $p$,                     $b(p)=\frac12(R_H+R_B)(p)$, $c(p)=\frac12(R_H-R_B)(p)$, and $a(p)=\frac{1+0.25U}{2(R_H+R_B)}(p)$, and $\hat{\theta}(p)$ is the wind direction with the upslope. Here, $U(p)$ is the speed of the wind in mid-flame, $R_H(p)$ is the downwind propagation speed, and $R_B(p)$ is the upwind propagation speed and all are determined by the data and tables \ref{t1} and \ref{t2} of Appendix \ref{apen3}.  The model is valid if $R_H^2(p)<\frac98(1+0.25U)(p)$.

Moreover, one finds the fire front at each time $T$ using each of the methods below.
\begin{itemize}
    \item[(i)] Finding the set $\{\gamma(T)\ |\ \gamma(t)\ \text{ are fire rays from } A\}$.       \item[(ii)] Applying Huygens' principle to $A$, where at each point $p\in A$, the spherical fire front is $\{\gamma(1)\}$, where $\gamma(t)$'s are solutions of equations system \eqref{geo-fin} such that $\gamma(0)=p$ and $\gamma'(0)=v$ satisfies the equation \ref{sph}. 
   \end{itemize}

\vspace{5mm}\textit{Validation of Model 4:} Using some  arguments analogous to those in {Models 2} and {3} one finds 
the Finsler metric that models the propagation as follows 
     \begin{equation}\label{fins4}
        F(v)=\frac{||v||}{\frac{a(p)b(p)}{\sqrt{(a^2(p)-b^2(p))\cos^2(\theta-\hat{\theta}(p))+b^2(p)}}+c(p)\cos(\theta-\hat{\theta}(p))}, 
    \end{equation}
    where $\theta$ is the direction of $v$ with the uphill. By the same discussions as {Models 2} and {3}, one finds the rays and waves.

\begin{remark}
   According to {Models 1-4}, to determine the fire front, we can either trace the fire rays or obtain spherical fire fronts and apply Huygens' principle. In both approaches, finding the Finsler geodesics is essential. For the fire rays, we focus on geodesics aligned with the propagation direction, while for spherical fire fronts, geodesics in all directions are needed.
 Both methods yield the same propagation model in simple cases, such as homogeneous fuel distribution or the absence of wind. However, when conditions vary significantly across space, the problem becomes more complex. In these cases, geodesics in the direction of propagation may no longer serve as a minimizer over long time intervals. To address this, it is advisable to avoid modeling fire fronts over extended periods. Instead, we should calculate spherical fire fronts for shorter intervals and apply Huygens' principle gradually.
\end{remark}

\subsection{Flowcharts}
We provide two flowcharts detailing the step-by-step process of finding the model to facilitate using {Models 1-4}. Flowchart \ref{apen1} models the propagation in space with homogeneous vegetation and environmental characteristics and Flowchart \ref{apen2} models the propagation in space with homogeneous and environmental vegetation characteristics. 
\subsubsection{Flowchart for finding  fire fronts for homogeneous vegetation characteristics}\label{apen1}
\begin{tikzpicture}[node distance=2cm]
\node (start) [startstop] {Start:Given fire front $A$};
\node (inp) [decision, right of=start, xshift=3cm] {wind blows?};
\node (ind1) [process, below of=inp, xshift=-3cm]{Calculate $R_0$ and $\phi_s$ by Appendix \ref{apen3} and data};
\node (ind2) [process, below of=inp, xshift=3cm] {Calculate $R_0, U, \phi_w, \phi_s, \hat{\theta}, R_B$ by Appendix \ref{apen3} and data};
\node (dat1) [decision, below of=ind1, yshift=0cm] {Need fire rays?};
\node (dat12) [decision, below of=ind2, yshift=0cm] {Need fire rays?};
\node (ph12) [process, below of=dat12, yshift=-1cm, xshift=0cm] { From each $p\in A,$ fire ray is $p+tv:v$ satisfies \eqref{con1} and \eqref{con2}};
\node (dat2) [process, below of=dat12, yshift=-1cm, xshift=4cm] {At each $p\in A,$ spherical fire front of radius $T$ is $p+Tv:v$ satisfies \eqref{con1}};
\node (sp1) [process, below of=dat1, yshift=-1cm, xshift=-2cm]{At each $p\in A,$ spherical fire front of radius $T$ is $p+Tv:v$ satisfies \eqref{alp1}};
\node (ph1) [process, below of=dat1, yshift=-1cm, xshift=2cm] {From each $p\in A$, fire ray is $p+tv:v$ satisfies \eqref{alp1} and \eqref{alp12}};
\node (wr1) [process, below of=ph1,  yshift=-0.5cm] {Fire front after $T$ unit time is $\{p+Tv:p\in A\}$};
\node (huy) [process, below of=sp1, yshift=-2cm, xshift=0cm] {Apply Huygens' principle on $A$ to find the fire front after $T$ time unit};
\node (huy2) [decision, below of=huy, xshift=4cm] {Need to continue?};
\node (back) [process, below of=huy2, xshift=0cm, yshift=0cm, text width=11cm] {Substitute $A$ with the updated fire front. If conditions have altered, return to the beginning; otherwise, backtrack four steps};
\node (stop) [startstop, right of=huy2, xshift=5cm, yshift=0cm] {Stop};
\draw [arrow] (start) -- (inp);
\draw [arrow] (ind1) -- (dat1);
\draw [arrow] (ind2) -- (dat12);
\draw [arrow] (dat12) --  node[anchor=west] {yes} (ph12);
\draw [arrow] (dat12) -|  node[anchor=south] {no} (dat2);
\draw [arrow] (dat1) -|  node[anchor=east] {no} (sp1);
\draw [arrow] (sp1) -- (huy);
\draw [arrow] (dat2) |- (huy);
\draw [arrow] (huy) |- (huy2);
\draw [arrow] (huy2) -- node[anchor=south] {no} (stop);
\draw [arrow] (ph1) -- (wr1);
\draw [arrow] (ph12) |- (wr1);
\draw [arrow] (inp) |- node[anchor=north] {no} (ind1);
\draw [arrow] (inp) -| node[anchor=west] {yes} (ind2);
\draw [arrow] (huy2) -- node[anchor=east] {yes} (back);
\draw [arrow] (dat1) -| node[anchor=west] {yes} (ph1);
\draw [arrow] (wr1) -- (huy2);
\end{tikzpicture}

\subsubsection{Flowchart for Determining Propagation in Nonhomogeneous Vegetation Characteristics}\label{apen2}
\begin{tikzpicture}[node distance=2cm]
\node (start) [startstop] {Start:Given fire front $A$};
\node (dat) [process, right of=start, xshift=4cm] {Choose a number of points $q$ of terrain around $A$};
\node (inp) [decision, below of=dat, yshift=.2cm, xshift=0cm] {wind blows?};
\node (ind1) [process, below of=inp, yshift=0cm, xshift=-3cm]{At each point \( q \), calculate \( R_0 \) and \( \phi_s \) using Appendix \ref{apen3} and the given data};
\node (ind3) [process, left of=ind1, xshift=-2cm]{Develop functions \(R_0\) and \(\phi_s\) that estimate the corresponding values of  \(R_0\) and \(\phi_s\) at each point \(q\).  Then, determine the metric \( F \) using equation \eqref{fins3}};
\node (ind2) [process, below of=inp, xshift=2cm, yshift=0cm] {At each point $q$, calculate $R_0, U, \phi_w, \phi_s, \hat{\theta}, R_B$ by Appendix \ref{apen3} and data};
\node (ind4) [process, right of=ind2, xshift=2cm]{Develop functions $R_0, U, \phi_w, \phi_s, \hat{\theta}, R_B$ that estimate the corresponding values of  $R_0, U, \phi_w, \phi_s, \hat{\theta}, R_B$ at each point \(q\).  Then, determine the metric \( F \) using equation \eqref{fins4}};
\node (dat1) [decision, below of=ind3, yshift=-1.5cm] {Need fire rays?};
\node (dat12) [decision, below of=ind4, yshift=-2cm] {Need fire rays?};
\node (ph12) [process, below of=dat12, yshift=-.5cm, xshift=0cm] {At each $p\in A$, find vectors $v$ satisfying \eqref{sph} and \eqref{sph2}};
\node (wr11) [process, below of=ph12,  yshift=-1cm] {Fire ray from $p\in A$ is solution of \eqref{geo-fin}: $\gamma(t)=(x(t),y(t))$ with $\gamma(0)=p$ and $\gamma'(0)=v$};
\node (dat2) [process, below of=dat12, yshift=2cm, xshift=-4cm] {At each $p\in A$, find spherical fire front as $\{\gamma(1)\}$: $\gamma(t)=(x(t),y(t))$ is solution of \eqref{geo-fin} with $\gamma(0)=p$ and $\gamma'(0)=v$ satisfying \eqref{sph}};
\node (sp1) [process, below of=dat1, yshift=-4cm, xshift=0cm]{At each $p\in A$, find spherical fire front as $\{\gamma(1)\}$: $\gamma(t)=(x(t),y(t))$ is solution of \eqref{geo-fin} with $\gamma(0)=p$ and $\gamma'(0)=v$ satisfying \eqref{alp2}};
\node (ph1) [process, right of=dat1, yshift=0cm, xshift=2.1cm] {At each $p\in A$, find vectors $v$ satisfying \eqref{alp2} and \eqref{alp3}};
\node (wr1) [process, below of=ph1,  yshift=-1cm, xshift=-.2cm] {Fire ray from $p\in A$ is solution of \eqref{geo-fin}: $\gamma(t)=(x(t),y(t))$ with $\gamma(0)=p$ and $\gamma'(0)=v$};
\node (wr2) [process, below of=wr1,  yshift=-1cm, xshift=5cm] {Fire front after $T$ unit of time is  $\{\gamma(T)\}$};
\node (huy) [process, below of=wr1, yshift=-1cm, xshift=.3cm] {Apply Huygens' principle on $A$ to find the fire front after $T$ time unit};
\node (huy2) [decision, below of=huy, xshift=2cm] {Need to continue?};
\node (back) [process, below of=huy2, xshift=-2cm, yshift=0cm, text width=11cm] {Substitute $A$ with the updated fire front. If conditions have altered, return to the beginning; otherwise, backtrack four steps};
\node (stop) [startstop, right of=huy2, xshift=3cm, yshift=-2cm] {Stop};
\draw [arrow] (start) -- (dat);
\draw [arrow] (dat) -- (inp);
\draw [arrow] (inp) -| node[anchor=south] {no} (ind1);
\draw [arrow] (inp) -| node[anchor=south] {yes} (ind2);
\draw [arrow] (ind1) -- (ind3);
\draw [arrow] (ind3) -- (dat1);
\draw [arrow] (ind2) -- (ind4);
\draw [arrow] (ind4) -- (dat12);
\draw [arrow] (dat12) --   node[anchor=west] {no}(ph12);
\draw [arrow] (dat12) --  node[anchor=south] {yes} (dat2);
\draw [arrow] (dat1) --  node[anchor=east] {no} (sp1);
\draw [arrow] (sp1) -- (huy);
\draw [arrow] (dat2) -- (huy);
\draw [arrow] (huy) -| (huy2);
\draw [arrow] (huy2) -| node[anchor=west] {no} (stop);
\draw [arrow] (ph1) -- (wr1);
\draw [arrow] (ph12) -- (wr11);
\draw [arrow] (huy2) -- node[anchor=east] {yes} (back);
\draw [arrow] (dat1) -- node[anchor=north] {yes} (ph1);
\draw [arrow] (wr1) -| (wr2);
\draw [arrow] (wr11) -- (wr2);
\draw [arrow] (wr2) -| (huy2);
\end{tikzpicture}

\section{Examples}

In Examples \ref{ex1} and \ref{ex2}, we simulate wildfire propagation scenarios on sloped terrain with smoothly distributed vegetation properties. Two distinct atmospheric configurations are considered: one without wind and one with wind. This dual approach enables a more precise analysis of fire behavior by assigning specific metrics to each vegetation type, thereby simplifying the analysis compared to applying a uniform metric across the entire terrain.

\begin{example}
    \label{ex1}

    This example demonstrates how small changes in conditions can lead to different propagation models. Figures~\ref{c1}-\ref{c3} illustrate these models, emphasizing the impact of subtle variations in vegetation or atmospheric characteristics. In Fig.~\ref{c1}, the data are \( R_0 = 1.8 - 0.6 \cos(x+y) \) and \( \phi_s = 0.45 \). As observed, the model demonstrates that the fire exhibits a more rapid propagation pattern toward the northeast direction, indicated by the elongated shape in this direction. Additionally, in certain regions of the figure, particularly where several wavefronts intersect, the behavior suggests complex interactions, likely due to variations in vegetation characteristics. These intersections may represent areas where differing propagation rates cause overlapping wavefronts, highlighting the model's sensitivity to heterogeneous conditions.

    In Figs.~\ref{c2}, the data are $\phi_s = 0.3$ and $R_0 = 3.5+\cos^2y$. Unlike the previous model, there are no intersections between wavefronts. This supports the idea that the propagation is steady and does not involve complex interactions or variations in the medium.
    
In Fig.~\ref{c3}, the fire spreads consistently during the first $10$ hours with $R_0 = 1.8 - \cos(x + y)$ and $\phi_s = 0.3$. During the next $20$ hours, the wind blows toward the northwest at a velocity of $U = 7, \text{km/h}$, with $R_0 = 3.3$, $\phi_s = 0.5$, and $\phi_w = 0.5$. As expected, the figure shows that the propagation speed is higher in the downwind direction, accompanied by lower fire intensity. The innermost contours are oval-shaped, further confirming the anisotropic influence of the source. This suggests that conditions near the ignition point already vary in their effect on propagation.

 Figs.~\ref{c1}-\ref{c3} particularly emphasize high-risk areas for rapid fire growth and potential fire trap formation.

\begin{figure}[h!]
			\centering
           \begin{subfigure}[b]{0.45\textwidth}
         \centering
         \includegraphics[width=.7\textwidth]{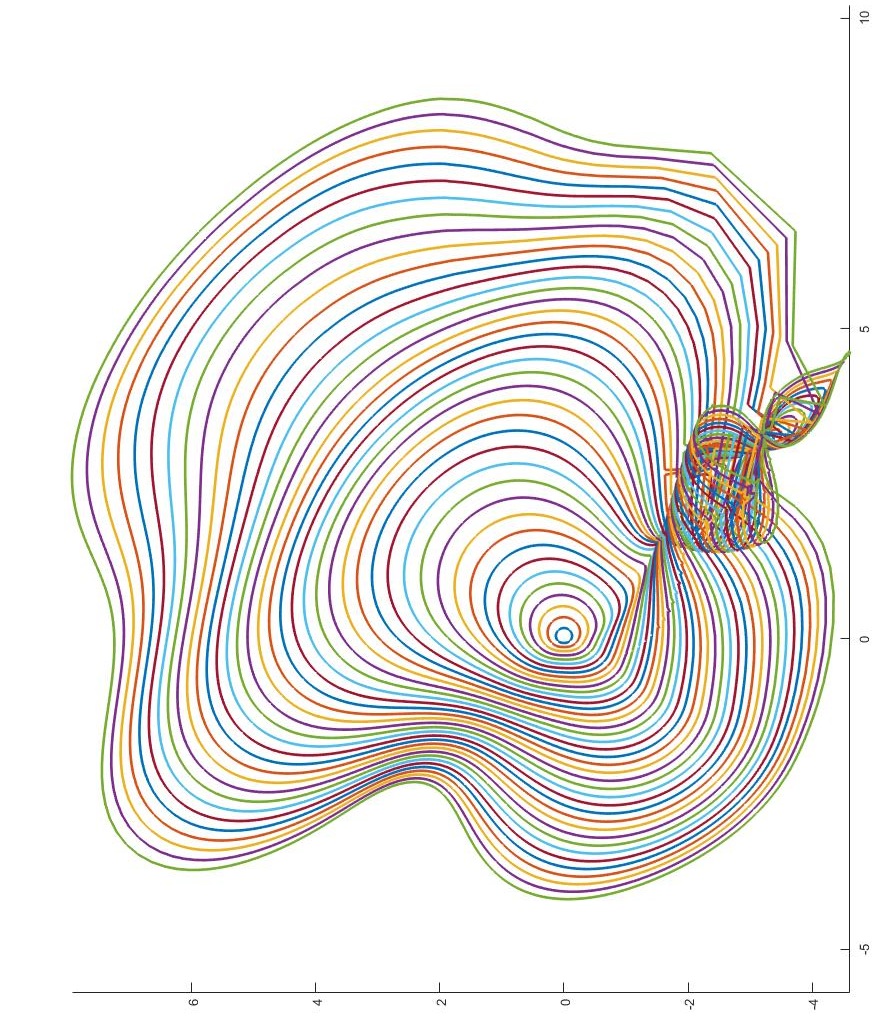}
         \subcaption{$R_0=1.8-0.6\cos(x+y), \phi_s=0.45$}\label{c1}
     \end{subfigure}
     \hfill
     \begin{subfigure}[b]{0.5\textwidth}
         \centering
         \includegraphics[width=.65\textwidth]{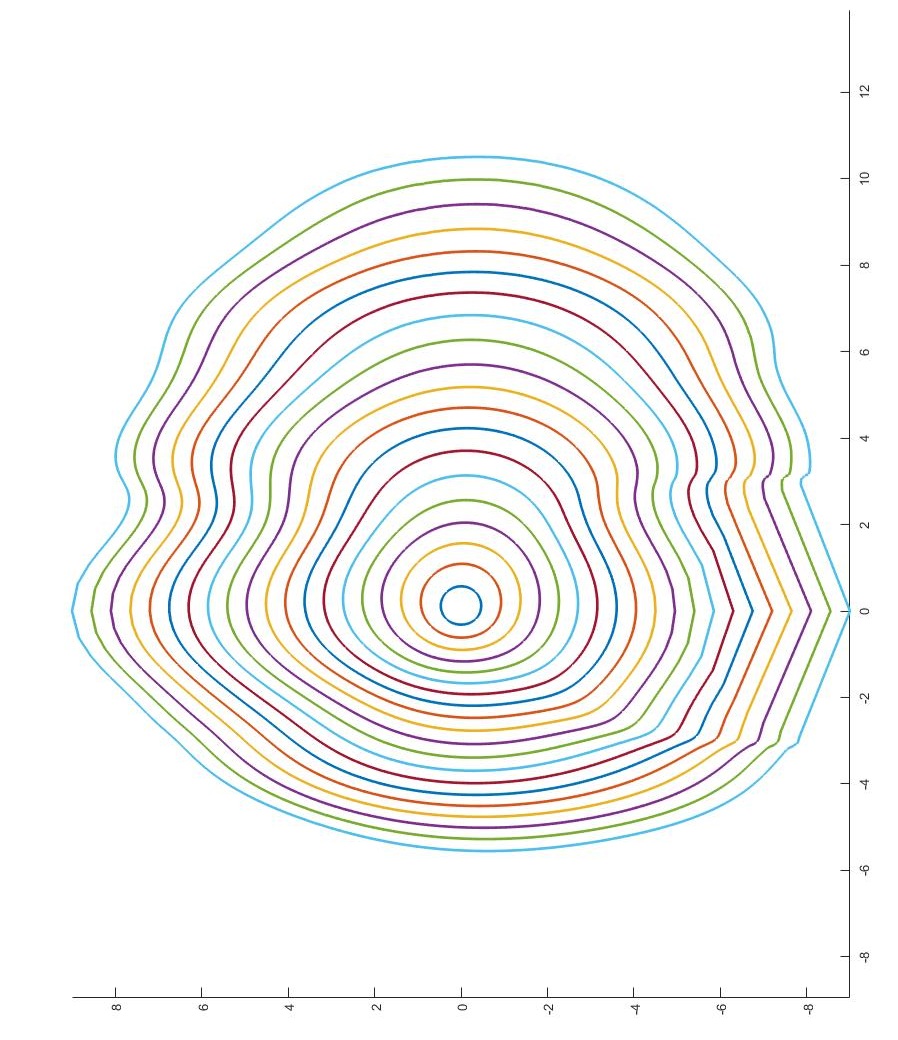}
         \subcaption{$R_0=3.5+\cos^2(y), \phi_s=0.3$}\label{c2}
     \end{subfigure}
     \centering
           \begin{subfigure}[b]{0.5\textwidth}
         \centering
         \includegraphics[width=.65\textwidth]{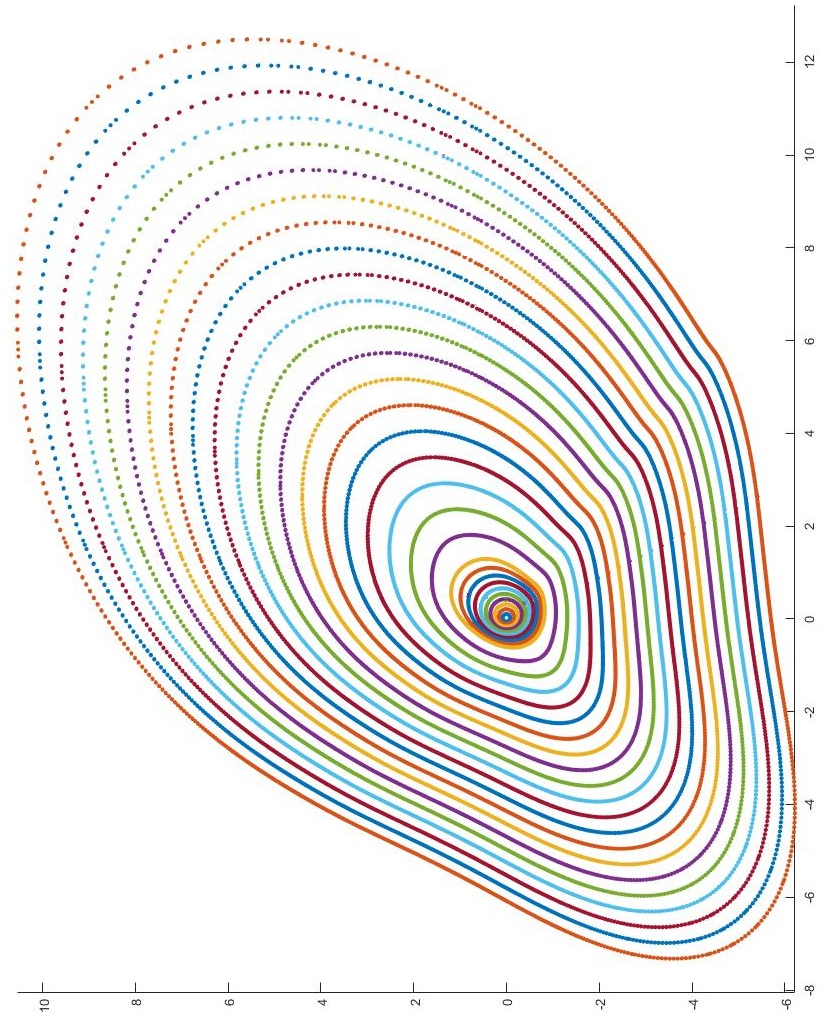}
         \subcaption{$R_0=1.8-\cos(x+y), \phi_s=0.3$ for the first $10$ hours of propagation and $U=7km/h, R_0=3.3, \phi_s=0.45,  \phi_w=0.5$, for the next $20$ hours of propagation.}\label{c3}
     \end{subfigure}
            \caption{Modeling the propagation with smooth distribution of vegetation and uniform environmental conditions.}\label{slope}
     \end{figure}
\end{example}

\begin{example}\label{ex2}
This example illustrates two methods proposed in this work for modeling propagation. 
    This example applies Huygens' principle (Fig.~\ref{r1}) and fire rays (Fig.~\ref{r2}) methods to model propagation with wind and varying vegetation conditions across two stages. In the first stage, there is no wind and vegetation varies across the space; and the rate of spread and slope factor is \(R_0 = 2.8 - 1.6 \cos(x + y)\) and \(\phi_s = 0.182\), respectively. In the second stage, the wind blows upslope at $7$ km/h and vegetation is uniformly distributed; and the rate of spread, slope factor, and wind factor are \(R_0 = 4.3\), \(\phi_s = 0.6\), and \(\phi_w = 0.3\), respectively. 
    
    While finding fire rays offers a more precise propagation model, in complex scenarios, the long geodesics may not be minimizer for a long time period. In such cases, Huygens' principle is applied, as it identifies short geodesics. However, the provided model by applying Huygens's principle may not provide information about fire paths, fire intensity, or trap formation.

It is recommended that both methods be applied whenever possible to create a more reliable model. According to Fig.~\ref{slope2}, the two methods yield similar results during the initial fire propagation stage. However, as the fire spreads southwest, the models diverge. It is considered the more reliable model since Fig.~\ref{r2} shows no disruption in rays directed southwest. Fig.~\ref{r2} also indicates that the rays toward the north are more dispersed, suggesting a less intense fire, while the northeast shows more concentrated rays, implying higher fire intensity. These interpretations serve as preliminary insights into the applications of this work, and further verification using real-world cases is necessary.
 In contrast, areas with fewer dots indicate lower fire intensity. Fig.~\ref{r2} further demonstrates variability in fire spread, illustrating random fire paths that provide additional insights into the complexity of wildfire behavior on sloped terrain.
\begin{figure}[h!]
			\centering
           \begin{subfigure}[b]{0.45\textwidth}
         \centering
         \includegraphics[width=.65\textwidth]{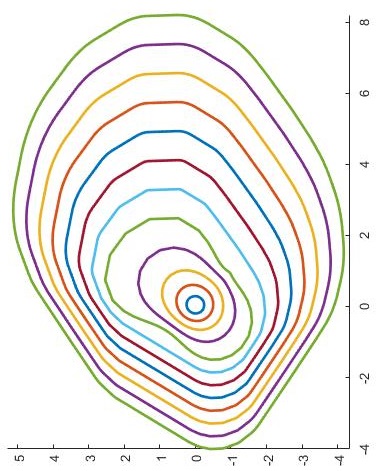}
         \subcaption{Fire model applying Huygens' principle.}\label{r1}
     \end{subfigure}
     \hfill
     \begin{subfigure}[b]{0.45\textwidth}
         \centering
         \includegraphics[width=.65\textwidth]{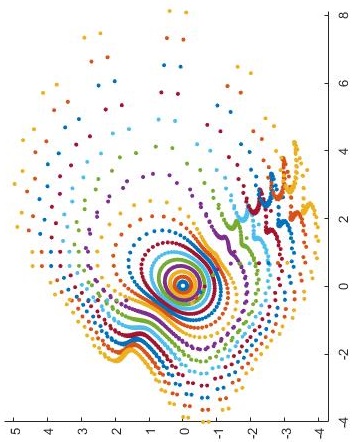}
         \subcaption{Fire model by finding the fire rays.}\label{r2}
     \end{subfigure}
           \caption{Comparing the methods of modeling using Huygens' principle and rays with $\Delta t=2$ hours.}\label{slope2}
     \end{figure}
 
\end{example}

\section{Conclusion and Final Remarks}
This work combined the Rothermel model and Huygens' principle with advanced geometric techniques to develop a systematic framework for understanding and predicting wildfire behavior in complex environments. We modeled wildfire propagation on an inclined plane, considering both scenarios windless and windy conditions, and provided equations of fire fronts and rays. By directly incorporating vegetation characteristics into these equations, our models offer a more accurate and practical representation of wildfire propagation.

We derived spherical fire front equations to apply Huygens' principle and approximate fire fronts, enabling simulators to model propagation based on geometric techniques.

For a more specific verification, we addressed propagations with both homogeneous and non-homogeneous vegetation characteristic distributions. Cases with homogeneous vegetation characteristics involve less complex calculations, reducing potential errors. These cases are particularly useful for laboratory studies and for analyzing real-world scenarios where vegetation characteristics can be considered homogeneous across the entire space or within specific subspaces, such as agricultural lands. We addressed these cases in Models 1 and 2. Models 3 and 4 investigated cases with non-homogeneous vegetation characteristic distributions. these cases are suitable for more detailed fire behavior modeling, where variations in fuel properties and environmental conditions across the space require a more accurate prediction of fire spread patterns. These scenarios are ideal for testing the sensitivity of fire propagation to small variations in terrain, wind, and vegetation, enabling a nuanced understanding of how wind or vegetation properties influence fire dynamics. Additionally, they are useful for validating models that simulate fire behavior under complex, real-world conditions. Some examples are simulated to demonstrate the practical applications of these results.

Inserting the vegetation properties directly into the formulas, simplifying input requirements, and providing explicit formulas and relations make our approach a valuable asset for advancing research in fire science and enhancing disaster management strategies.

By applying the results of this work, one can identify the locations of fire fronts and fire paths. These insights offer several benefits, including:

\begin{itemize}
    \item[1.] Better emergency response planning through the better allocation of resources.
    \item[2.]  Risk assessment and management by taking preventative measures, such as creating firebreaks or clearing vegetation.
\item[3.]  Urban and regional planning by incorporating safety zones and buffer areas to protect critical assets.
\item[4.] Scientific research by improving fire behavior models and developing new fire management techniques, especially when applying simulators such as FARSITE.
\end{itemize}

\section*{Acknowledgment}
The author acknowledges the financial support of FAPESP (Grant
2022/15371-3).
\newpage
\section{Appendix: The formulas to calculate the frame}\label{apen3}
\begin{table}[h!]
\centering
\caption{Equations for the frame fire spread model \cite{andrews2018rothermel}}\label{t1}
\begin{tabular}{@{}lll@{}}
\toprule
Element & Equation & \\ \midrule
Backing fire spread rate&$R_{B} = R_{H} (\frac{1-e}{1+e})$&\\
 &$e = \frac{\sqrt{z^2-1}}{z}$&\\
 &$z = 1 + 0.25U$&\\
 {\small Spread rate in maximum spread direction (ft/min)} & $R_{H} = R_{0} (1+ \phi_w+\phi_s)$ & \\ 
No-wind, no-slope propagating flux (Btu/ft2/min)&$R_{o} = I_{R} \xi$&\\
 Reaction intensity (Btu/ft\(^2\)-min)&$I_R = \Gamma' w_n h \eta_M \eta_S$&\\
Optimum reaction velocity (min) & $\Gamma' = \Gamma'_{\max} \left( \frac{\beta}{\beta_{op}} \right)^A \exp \left[ A \left( 1 - \frac{\beta}{\beta_{op}} \right) \right]$ &\\ 
 & $A = 133 \sigma ^{-0.7913}$ & \\
Maximum reaction velocity (min) & 
$\Gamma'_{max} = \sigma ^{1.5} ( 495 + 0.0594 \sigma^{1.5})^{-1}$ & \\ 
Wind factor & 
$\phi_w = C U^B \left( \frac{\beta}{\beta_{op}} \right)^{-E}$ & \\ 
Slope factor & 
$\phi_s = 5.275 \beta^{-0.3} (\tan \phi)^2$ &\\ 
Optimum packing ratio & 
$\beta_{op} = 3.348 \sigma^{-0.8189}$ &  \\ 
Packing ratio & 
$\beta = \rho_b / \rho_p$ & \\ 
Oven-dry bulk density (lb/ft\textsuperscript{3}) & 
$\rho_b = w_o / \delta$ &  \\ 
Net fuel load (lb/ft\textsuperscript{2}) & 
$w_n = w_o (1 - S_T)$ &  \\
Moisture damping coefficient & 
$\eta_M = 1 - 2.59 r_M + 5.11 (r_M)^2 - 3.52 (r_M)^3$ &\\ 
 & $r_M = M_f / M_x$ (max = 1.0) & \\ 
Mineral damping coefficient & 
$\eta_s = 0.174 S_e^{-0.19}$ (max = 1.0) & \\
Propagating flux ratio & 
$\xi = \frac{exp{[(0.792 + 0.681 \sigma^{0.5})(\beta + 0.1)]}}{192 + 0.2595 \sigma}$ & \\ 
 & $C = 7.47 exp(-0.133 \sigma^{0.55})$ & \\ 
 & $B = 0.02526 \beta^{0.54}$ & \\ 
 & $E = 0.715 \exp \left[ -3.59 \times 10^{-4} \delta \right]$ & \\ 
Effective heating number & 
$\epsilon = \exp (-138 / \delta)$ &  \\ 
Heat of preignition (Btu/lb) & 
$Q_{ig} = 250 + 1116 M_x$ &\\ \bottomrule
\end{tabular}
\end{table}

\begin{table}[h]
    \centering
    \caption{Parameter Definitions and Notes \cite{andrews2018rothermel}}\label{t2}
\begin{tabular}{@{}lll@{}}
\toprule
Element & Equation & \\ \midrule        
        \textbf{Symbol} & \textbf{Parameter} & \textbf{Notes} \\ 
        $h$ & Low heat content (Btu/lb) & Often 8,000 Btu/lb \\ 
        $S_T$ & Total mineral content (fraction) & Generally 0.0555 \\ 
        $S_e$ & Effective mineral content (fraction) & Generally 0.010 \\ 
        $\rho_p$ & Oven-dry particle density (lb/ft\(^3\)) & Generally 32 lb/ft\(^3\) \\ 
        $\sigma$ & Surface-area-to-volume ratio (ft\(^2\)/ft\(^3\)) & \\ 
        $w_0$ & Oven-dry fuel load (lb/ft\(^2\)) & Mean fuel array value \\ 
        $\delta$ & Fuel bed depth (ft) & \\ 
        $M_x$ & Dead fuel moisture of extinction (fraction) & \\ 
        $M_f$ & Moisture content (fraction) & Dry weight basis  \\ 
        $U$ & Wind velocity at midflame height (ft/min) & \\ 
        $\tan \phi$ & Slope steepness&  \\  \bottomrule
    \end{tabular}
  \end{table}
  
\newpage

\bibliography{bib}
\bibliographystyle{unsrt}

\end{document}